# PREDICTIVE AND STATISTICAL ANALYSES FOR ACADEMIC ADVISORY SUPPORT

Mohammed AL-Sarem

Department of Information Science, Taibah University, Al-Madinah Al-Monawarah, Kingdom of Saudi Arabia
mohsarem@gmail.com

*ABSTRACT*

*The ability to recognize students' weakness and solving any problem may confront them in timely fashion is always a target of all educational institutions. This study was designed to explore how can predictive and statistical analysis support the academic advisor's work mainly in analysis students' progress. The sample consisted of a total of 249 undergraduate students; 46% of them were Female and 54% Male. A one-way analysis of variance (ANOVA) and t-test were conducted to analysis if there was different behaviour in registering courses. Predictive data mining is used for support advisor in decision making. Several classification techniques with 10-fold Cross-validation were applied. Among of them, C4.5 constitutes the best agreement among the finding results.*

## 1. INTRODUCTION

The ability to recognize students' weakness and solving any problem may confront them in timely fashion is always a target of all educational institutions. At this end, colleges and universities began to implement so-called academic advising affairs. According to National Academic Advising Association[1], the academic advising is "a series of intentional interactions with a curriculum, pedagogy, and a set of student learning outcomes". At a time not so long ago, students were responsible for their own choices and the faculty advisor had primarily become assisting students with the transition from high school to college [1]. However, nowadays, situation is extended to include guiding students to select courses to register in each semester and fulfil the degree requirement.

In the Taibah University (TU), many tasks rely on the faculty advisor. He serves either as a coordinator of learning experiences through course, facilitator of communication, career planning and academic progress review, or as supporter students while they navigate the curriculum plan. So, the main task for the academic advisor is to help students in selection courses at each semester that guarantees complete the degree requirements in a timely manner. Each semester student should to meet his advisor a face-to-face. Often, the advising process is manual even if there is some tool to access students' transcription. However, this process has many drawbacks: labour intensive, time consumption, human advisors, limited knowledge and the large number of students compared to the number of advisors [2][3][4]. To cope with these problems, several approaches have been proposed. Henning in [5] discussed problems of the manual advising process such as limited number of advisors, advisors availability, problem of incompetent advisors, as well as, the serious consequences may occur if mistakes are made, for example, graduation delay, major or college drop out. In [3], authors proposed a CBR advising system that can be used for automating the manual process of academic advising. Another attempt to build an advising system is the rule-based advisory systems proposed by Al Ahmar [6] that aims to assist the students in selecting courses each semester. Both the students and advisors were provided with a useful tool for quick and easy course selection and evaluation of various alternatives.

---
[1] http://www.nacada.ksu.edu/Clearinghouse/AdvisingIssues/Concept-Advising.htm

Business intelligence tools could be very useful for that purpose. According to [7], applying statistical analyses and data mining algorithms on collected data can enable further understanding and improvements. At this end, we found, in literature, several attempts to design computer applications. These applications are categorized based upon data mining objectives into two categories: a description objective and a prediction objective [8]. In [9], authors proposed an advising system to help undergraduate students during the registration period. The proposed system uses a real data from the registration pool, then applies association rules algorithm to help both students and advisors in selecting and prioritizing courses.

At this end, the goals of this paper are twofold: 1) to show how the data mining techniques can support the advisor's work mainly in analysis students' progress and also in scheduling their plan, 2) and, to present some recommendation that improve academic advising at TU.

This paper is organized as follows: in section 2, academic advising process in TU is presented. Section 3 introduces the data mining techniques and used tools. Section 4 presents the experiment methodology including available data and experiment design. Section 5 presents some predictive and statistical analyses that can be helpful for improving the academic advisory process. Finally, section 6, concludes the paper.

## 2. ACADEMIC ADVISING AFFAIR AT TAIBAH UNIVERSITY

Taibah University[2] was established in 2003 by the royal decree under the number 22042 that integrates both the Muhammad bin Saud University and King Abdulaziz University located in Medina into one independent university sited in Medina. Currently, TU includes 22 colleges. The college of computer science and engineering (CCSE) includes three departments: the computer science (CS), information science (IS), and computer engineering (CE). Statistics in Table 1show percentage of students to academic advisors year by year[3].

Although there is an effort to increase amount of academic advisors, the advisory problems still not resolved completely. Next subsection presents the advisory procedure in TU and duties of academic advisors.

Table 1: Students number per year at TU

| Year | | Department | | | Total number of students | Number of academic advisor | Percentage of students to academic advisor |
|---|---|---|---|---|---|---|---|
| | | SC | IS | CE | | | |
| 2003/2008 | Boys | 319 | 85 | 62 | 1060 | 29 | 36:1 |
| | Girls | 445 | 104 | 51 | | | |

---

[2] https://www.taibahu.edu.sa/Pages/EN/Home.aspx
[3] https://www.taibahu.edu.sa/Pages/AR/Sector/SectorPage.aspx?ID=56&PageId=6

| | | | | | | | |
|---|---|---|---|---|---|---|---|
| 2008/2009 | Boys | 318 | 102 | - | 1020 | 30 | 34:1 |
| | Girls | 431 | 169 | - | | | |
| 2009/2010 | Boys | 479 | 121 | - | 1275 | 31 | 41:1 |
| | Girls | 419 | 256 | - | | | |
| 2010/2011 | Boys | 479 | 121 | - | 1276 | 31 | 41:1 |
| | Girls | 419 | 256 | - | | | |
| 2011/2012 | Boys | 456 | 142 | - | 1130 | 37 | 31:1 |
| | Girls | 356 | 176 | - | | | |
| 2013/2014 | Boys | 221 | 167 | 113 | 1218 | 42 | 29:1 |
| | Girls | 336 | 242 | 139 | | | |
| 2014/2015 | Boys | 230 | 179 | 109 | 1234 | 47 | 26:1 |
| | Girls | 208 | 376 | 142 | | | |

## DUTIES OF ACADEMIC ADVISOR

Faculty academic advising has a significant impact on a student's academic success [10]. This work is dependent on both academic advisor and academic programs of educational institutions. In TU, the academic programs follow the credit hours system. The academic year is divided into three semesters; two of them are regular (spring and autumn semester) and summer semester for students who failed the regular semesters. The maximum hours that students can take each regular semester are 19 hours (except graduate student whom allowed to take extra three hours), while only six hours are allowed at summer semester. The minimum credit hours are 13 each semester. For students who failed any semester are forced to take the minimum credit hours at the next semester. At beginning of the academic advising process, all students are distributed according their specialization into several academic advisors (often, a member of academic stuff). According to academic advising guideline in TU, the academic advisor is responsible for doing the following:

1. Help students in adaptation with specialization, especially freshmen, and work hardly to overcome the obstacles and problems.

2. Follow-up to the level of students each semester, encourage and draw a good study plan that ensures the improvement their educational level.

3. View the study plans of the department for students who supervises them and know the courses for each level.

4. Determine which courses that may delay student's graduation at the specified time

5. Help students to correctly register their plan of study according to the rules of acceptance and registration deanship.

6. Create an e-portfolio including student name, current student transcription, current plan of study, and department plan of study.

7. Write a report at the end of courses registration, and summarize the problems that confronted students during semester (see Fig.1).

| First Academic Semester | | Second Academic Semester | | Summer Semester | |
|---|---|---|---|---|---|
| Recommended Courses | Actually selected Courses | Recommended Courses | Actually selected Courses | Recommended Courses | Actually selected Courses |
| 1. | 1. | 1. | 1. | 1. | 1. |
| 2. | 2. | 2. | 2. | 2. | 2. |
| 3. | 3. | 3. | 3. | 3. | 3. |
| 4. | 4. | 4. | 4. | 4. | 4. |
| 5. | 5. | 5. | 5. | 5. | 5. |
| 6. | 6. | 6. | 6. | 6. | 6. |
| Is there any problem confronted student? Yes ☐ No ☐ | | Is there any problem confronted student? Yes ☐ No ☐ | | Is there any problem confronted student? Yes ☐ No ☐ | |
| Problem type: Academic ☐ Psychological ☐ Social ☐ | | Problem type: Academic ☐ Psychological ☐ Social ☐ | | Problem type: Academic ☐ Psychological ☐ Social ☐ | |
| Solution: ................................ ................................ ................................ | | Solution: ................................ ................................ ................................ | | Solution: ................................ ................................ ................................ | |
| Student's signature: Advisor's name: Advisor's signature: | | Student's signature: Advisor's name: Advisor's signature: | | Student's signature: Advisor's name: Advisor's signature: | |

Figure 1: Annual Academic Report using at TU

## 3. APPROPRIATE DATA MINING TECHNIQUES AND THE USED TOOLS

Data mining is a rising area of research and development, both in academic world as well as in business [11]. Data mining is the combination of statistical modelling, database storage, information science, visualization, and artificial intelligence.

Applying data mining algorithm in academic world produces new academic research area known today as "educational data mining" (EDM). According to educational data mining association[4], the EDM is "an emerging discipline, concerned with developing methods for exploring the unique types of data that come from educational settings, and using those methods to better understand students, and the settings which they learn in". Romero & Ventura, in [12], reviewed hundreds of EDM articles and proposed desired EDM objectives based on the roles of users. Currents works use different techniques including: association, classification, clustering, prediction and sequential patterns.

Classification is most frequent task in Data mining. Its main goals are to classify objects into a number of categories referred to as classes. Figure 2 shows a simplified model of classification task in data mining. To evaluate and compare learning algorithms, several statistical methods are used. Among of them, the cross-validation is common one. In cross-validation, data is divided into two sets: the first is training set which is used to learn or train model, and the second one is test set to validate the model.

There are several forms of cross-validation: k-fold cross-validation, re-substitution validation, hold-out validation, leave-one-out cross-validation and repeated k-fold cross-validation. Among of them, the k-fold cross-validation is the basic [13] in which the data is partitioned into **k** equally (or nearly equally) folds, then during the training and validation process, a different fold of data is held-out for validation while the remaining $k - 1$ folds are used for training. The most common cross-validation types, especially in data mining and machine learning, is 10-fold cross-validation since it tends to provide less biased estimation of the accuracy [14].

---
[4] http://educationaldatamining.org/

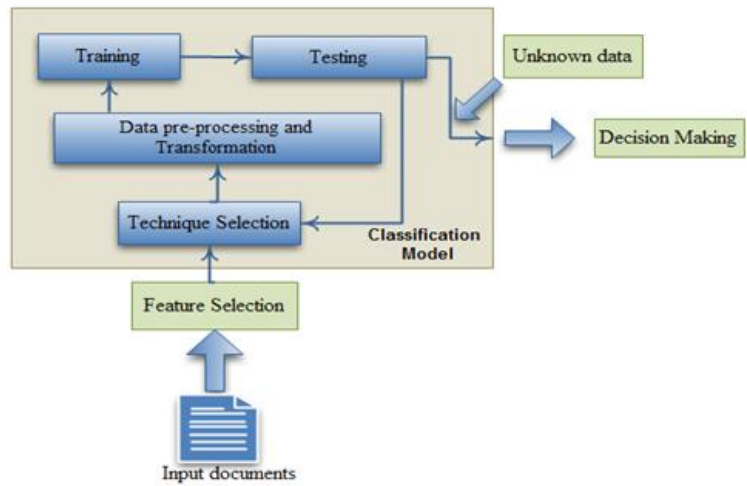

Figure 2: A simplified Classification Model

Below, current section gives briefly overview of the most common classification techniques. These techniques can be classified into three main groups: induction-based algorithms (decision tree induction), probability-based algorithms, and analogy-based algorithms.

### 3.1 DECISION TREE INDUCTION: C4.5 ALGORITHM

Algorithms of such group classify instances by sorting them based on feature values [15] in order to expose the structural information contained in the data [16]. To build the decision tree, a greedy algorithm constructs tree in a top-down recursive divide-and-conquer manner as follows:

1. for a given set $S$ of cases, a single attribute with two or more outcomes is selected as the root of the tree with one branch for each outcome (subsets $S_1, S_2,...,S_n$ ).

2. for all the cases belong to the same class or if $S$ is quite small, the tree is a leaf labelled with the most frequent class in $S$.

3. Apply the same procedure ( step 1, and 2) to each subset until it has the same class.

For finding the root of the tree (step 1), numerous methods are used: information gain, Gain ratio, Gini index, Chi square, G-statistics, Minimum Description Length (MDL) and Multivariate split [16]. Among the existing classification algorithms, the C4.5 [17] deserves a special mention due to its accuracy and performance [18]. Regarding the partition method, the C4.5 use two heuristic criteria to rank possible outcomes: information gain for minimizing the total entropy of the subsets $S_i$, and gain ratio [19].

### 3.2 PROBABILITY-BASED ALGORITHMS: NAÏVEBAYES CLASSIFIER

Algorithms of such group classify instances based on their probability distribution. The Naive Bayes classifier is a fast, easy to implement, relatively effective algorithm [20]. Furthermore, it does not require any complicated iterative parameter estimation schemes which makes it suitable for applying on huge data sets [21]. Simply, the algorithm predicts the class by computing conditional probability that a given sample belongs to a particular class. The Naïve Bayes Classifier is based on Bayes' Theorem:

$$P(i|x) = \frac{P(x|i)P(i)}{P(x)} \quad (1)$$

where,

**P(x|i)** - the conditional distribution of **x** for class **i** objects

**P(i)** - the – prior probability that an object will belong to class **i**.

So, to predict class of objects, the algorithm uses the following formula and compares the probabilities:

$$\frac{P(i)}{P(j)} = \frac{P(x|i)P(i)}{P(x|j)P(j)} \quad (2)$$

To estimate **P(i)**, we assume that the training set is a random sample from the overall population and **P(i)** is proportion of class **i** objects in the training set. On the other hand, to estimate **P(x|i)** we assume that objects of **x** are independent. So, $P(x|i) = \prod_{j=1}^{p} f(x_j|i)$. Knowing that, the ratio in Equation (2) becomes:

$$\frac{P(i)}{P(j)} = \frac{\prod_{k=1}^{p} f(x_k|i)P(i)}{\prod_{k=1}^{p} f(x_k|j)P(j)} = \frac{P(i)}{P(j)} \prod_{k}^{p} \frac{f(x_k|i)}{f(x_k|j)} \quad (3)$$

## 3..3 K-NEAREST NEIGHBOUR

The k-nearest neighbour algorithm is well-known classifier which is based on so-called learning by Analogy [15] where a group of **k** samples in training set are found and matched to the closest on the test sample. Often, the closest sample is defined by simple majority voting or by computing Euclidean distance between two points $X = \{x_1, x_2, \ldots, x_n\}$ and $Y = \{y_1, y_2, \ldots, y_n\}$ as follows:

$$d(X, Y) = \sqrt{\sum_{i=1}^{n}(X_i - Y_i)^2} \quad (4)$$

Basically, in a two-dimensional feature space with two target classes and number of neighbour **k**=5 (see Figure 3), the decision for $q_1$ is easy since all five of its nearest neighbours are of the same class which means it also can be classified as its neighbours. However, for $q_2$ the Euclidean distance should be computed firstly to decide to which class it belong. Generally said, the K-nearest neighbour algorithm has two main stages: the first is determination the nearest neighbour, and the second is classification the object using class of its nearest neighbour.

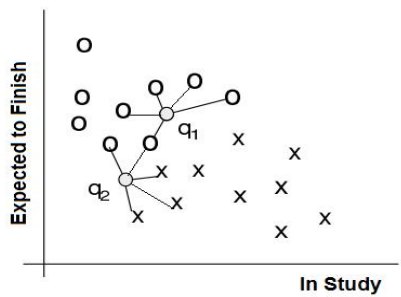

Figure 3: A simple example of 5-Nearest Neighbour Classification

# 4. EXPERIMENT METHODOLOGY

## 4.1 AVAILABLE DATA

Early recognition of students who are likely need more attention during his study is in the heart of academic advisory. The available data for current research was collected from undergraduate students of Information Science department at TU in Al-Madinah Al-Munawarah using the existing academic system. The collected data was of three academic programs (old, new and developed program). All available attributes are shown in Table 2.

Table 2: Available Attributes

| Attribute | Abbreviation | Attribute type | Availability By |
|---|---|---|---|
| Student Id | Sid | Numeric | System |
| Total Registered Credit Hours | Total_Reg_C_H | | |
| Total Gained Credit Hours | Total_Gain_C_H | | |
| Total Credit Hours. in the Current Semester | Total_Cur_C_H | | |
| Cumulated Grade Point Average | CUM. GPA | | Calculated by System |
| Different between Gained and Registered Credit hours | Diff_G_R_C_H | | Calculated manual |
| Learning Status | L_STATUS | Nominal | System |
| Gander | GEN | | |
| Advisory status | Ad_STATUS | | Advisor |
| Academic Plan of Study | Plan_Study | | |

The "Sid" attribute, here, included in order to help advisor in determination which academic plan of study was followed by student. However, "Ad_STATUS" was determined by experienced advisor. The "Total_Reg_C_H" refers to the registered hours in the system even if student decides to withdraw or postpone a course from the current semester. The "Diff_G_R_C_H" shows the different between the registered and gained hours.

## 4.2 EXPERIMENT DESIGN

Since the goal of this paper is to find a way to support academic advisory process, descriptive and predictive analyses were selected. The role of descriptive analysis is to give general insights over data, whereas predictive analysis supports advisor in decision making. Summarisation and visualisation are most appropriate techniques of descriptive data mining, while classification techniques are most common techniques of predictive analysis since they present easy and interpretable outputs.

Regarding the used tool, as the data come from various different sources, several steps were conduit to transform it into the appropriate format (see Fig. 4).

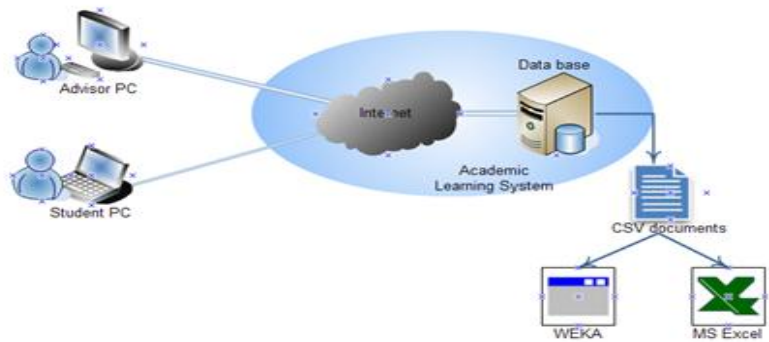

Figure 4: Transformation Procedure

For statistical analysis and data mining, both Weka[5] 3.0 and MS Excel 2013 with data analysis plug-in were used. The reason for this selection is that the Weka is the oldest and most successful open source data mining library and software which was later integrated as libraries in RapidMiner and R [22]. According to KDnuggets annual software poll [23], MS Excel holds the third place among most used data analysis tools.

## 5. ANALYSIS AND RESULTS

### 5.1 STATISTICAL ANALYSIS OF DATA

For conducting statistical analysis, in the beginning we used "L_STATUS" attribute for distinguishing between students who are expected to finished their learning from those are in study since students in the last years have better indicators of "Total_Reg_C_H" and "Total_Gain_C_H". Table 3 shows the related statistical analysis.

Table 3: Statistical analysis

| Student Group | | Total_Reg_C_H | Total_Gain_C_H | $\Delta(Total_{Reg_{C_H}} - Total_{Gain_{C_H}})$ | CUM. GPA |
|---|---|---|---|---|---|
| Mean µ | Expected | 175.5385 | 155.4872 | 20.05128 | 3.81359 |
| | In Study | 109.3381 | 85.09524 | 24.24286 | 3.370762 |
| St. dev. | Expected | 20.50911 | 21.13599 | 15.59175 | 0.581158 |
| | In Study | 41.40297 | 41.08295 | 12.7697 | 0.767264 |
| Median | Expected | 174 | 152 | 17 | 3.77 |
| | In Study | 103 | 80 | 22 | 3.33 |
| Mode | Expected | 164 | 155 | 16 | 3.75 |
| | In Study | 61 | 45 | 19 | 4.01 |
| Standard Error | Expected | 3.284085 | 3.384467 | 2.496678 | 0.09306 |
| | In Study | 2.857076 | 2.834993 | 0.881193 | 0.052946 |
| Kurtosis | Expected | 4.652146 | 7.2624 | 0.143323 | -0.26605 |
| | In Study | -0.19575 | 0.756821 | 3.596345 | -0.24391 |
| Skewness | Expected | -0.52991 | 2.553589 | 0.135872 | -0.17302 |
| | In Study | 0.418086 | 0.888382 | 0.268973 | -0.11597 |

---
[5] http://www.cs.waikato.ac.nz/ml/weka/downloading.html

Out of 249 students, there were a total of 39 (≈16%) expected to complete their study, 46% of them were Female and 54% Male. One-way ANOVA was also used to test differences whether the different between the registered and gained credit hours $\Delta(\text{Total}_{\text{Reg}_{C_H}} - \text{Total}_{\text{Gain}_{C_H}})$ between two groups were different. The test was significant at $p < 0.05$. Results show that really there was a different between both groups ($p_{sta} = 0.068789 > p$).

Table 4: One-way ANOVA test

| Source of Variation | SS | Df | MS | F | P-value | F crit. |
|---|---|---|---|---|---|---|
| Between Groups | 730.6691 | 1 | 730.6691 | 3.513325 | 0.068789 | 4.105456 |
| Within Groups | 7694.921 | 37 | 207.9708 | | | |
| Total | 8425.59 | 38 | | | | |

We refer this difference to that, on one hand, boys group did not receive enough advising during their study since they select courses without carefully. On the other hand, the "CUM. GPA" of boys group is less than the girls' which enforces the previous observation. Particularly, this leads us to the next question: Is there different between students with high "CUM. GPA" [3.76-5.00] and those with "CUM. GPA" [2.00-3.75] regarding the " Diff_G_R_C_H" attribute?

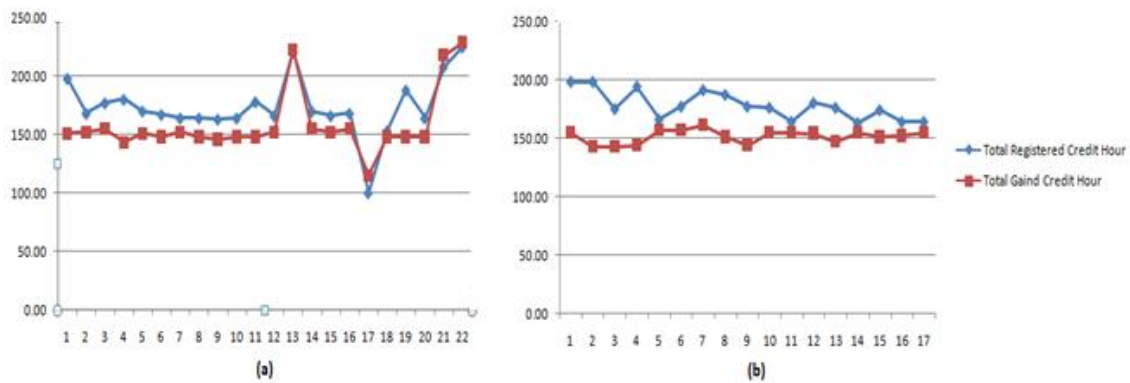

Figure 5:(a) - Difference between "Total_Reg_C_H" and "Total_Gain_C_H" for students with high "CUM. GPA", (b) with poor ""CUM. GPA"

Table 5: t-Test- Two-Sample Assuming Equal Variances

| | Group with high GPA (Good Group) | Group with Low GPA (Poor Group) |
|---|---|---|
| Mean | 19.05882353 | 26.17647059 |
| Variance | 107.8088235 | 210.7794118 |
| Observations | 17 | 17 |
| Pooled Variance | 159.2941176 | |
| Hypothesized Mean Difference | 0 | |
| Df | 32 | |
| t Stat | -1.644167436 | |
| P(T<=t) one-tail | 0.054966019 | |
| t Critical one-tail | 1.693888703 | |

| P(T<=t) two-tail | 0.109932039 |
| t Critical two-tail | 2.036933334 |

The t-test, in Table 4, reveals that there were significant differences at the ($p < 0.05$) level. Analysis also shows that students with poor GPA scored a group mean of (26.17) which is greater than the mean score of students with good GPA. We find this logically since poor students either re-take the courses in which they failed or they withdraw or postpone an already registered course taught by a strict instructor from the semester they are (Fig. 6).

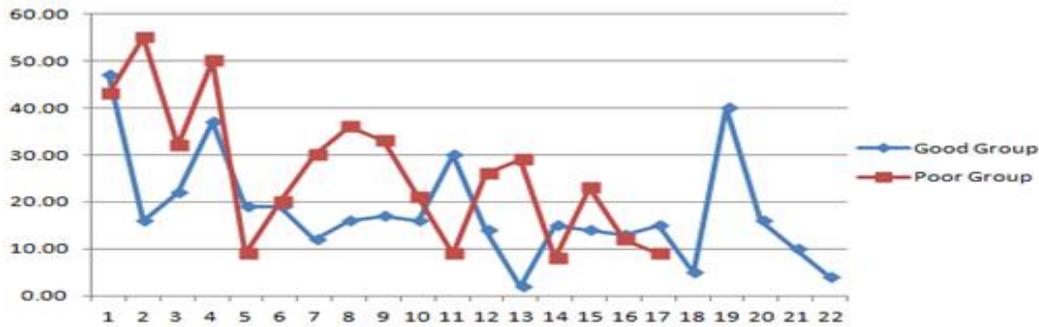

Figure 6: "Diff_G_R_C_H" between students with Good and Poor "CUM. GPA"

## 5.2 PREDICTIVE ANALYSIS

As mentioned previously, classification techniques are the most common techniques of predictive analysis that present easy and interpretable outputs. Armed with the algorithms known in machine learning field, we could determine those students who need more attention in more precise way.

For this purpose, the classification model was formed on the bases of whole gathered data (L_STATUS= "In study" or "expected to graduate"). Regarding the used algorithm, the C4.5, K-nearest neighbour [24], and NaïveBayes classifier [25] with 10-fold Cross-validation have been applied.

The accuracy of classifier is defined in terms of percentage of correct classified instances. Table 6 shows the obtained results after applying the previous algorithms.

Table 6: Related Statistical Analysis of the Used Algorithms

|  | correctly classified instances % | Kappa statistic | Mean absolute error | Root mean squared error | Relative absolute error % | Root relative squared error % |
|---|---|---|---|---|---|---|
| **C4.5** | 87.5502 % | 0.5461 | 0.1277 | 0.2751 | 61.339 % | 85.8932 % |
| **NaïveBayes classifier** | 87.9518 % | 0.5123 | 0.143 | 0.2678 | 68.692 % | 83.6293 % |
| **K-nearest neighbour** | 86.3454 % | 0.5126 | 0.1007 | 0.2912 | 48.399 % | 90.926 % |

From Table 6, it is apparent that NaïveBayes classifier gives the best results among the used algorithms, however, regarding the Kappa statistic [26], C4.5 constitutes the best agreement among the finding results. Furthermore, since the academic advisory essentially aims to help students with high risk or near to fail a semester, the C4.5 algorithm gives the best F-Measure (see Table 7).

Table 7: Algorithms Measurements

|  | Precision | Recall | F-Measure | Class |
|---|---|---|---|---|
| **C4.5** | 0.903 | 0.956 | 0.929 | Normal |
|  | 0.583 | 0.4 | 0.475 | Near To Risk |
|  | 1 | 0.9 | 0.947 | Under Risk |
|  | 0.862 | 0.876 | 0.866 | Weighted Avg. |
| **NaïveBayes classifier** | 0.889 | 0.985 | 0.935 | Normal |
|  | 0.714 | 0.286 | 0.408 | Near To Risk |
|  | 0.889 | 0.8 | 0.842 | Under Risk |
|  | 0.865 | 0.88 | 0.857 | Weighted Avg. |
| **K-nearest neighbour** | 0.897 | 0.941 | 0.919 | Normal |
|  | 0.52 | 0.371 | 0.433 | Near To Risk |
|  | 1 | 1 | 1 | Under Risk |
|  | 0.848 | 0.863 | 0.854 | Weighted Avg. |

The decision tree resulted from the C4.5 algorithm presents an interesting rule: "Students with highly difference between the registered credit hour "Total_Reg_C_H" and the gained credit hour "Total_Gain_C_H" are more closer to fail in a semester or more likely to have low GPA.

## 6. CONCLUSION

This paper presented a case study showing how can predictive and statistical analyses be helpful in academic advisory at the Information System Department, Faculty of Computer Science and Engineering, Taibah University. The sample of research consisted of a total of 249 undergraduate students; 46% of them were Female and 54% Male. A one-way analysis of variance (ANOVA) and t-test were conducted to analysis if there was different behaviour in registering courses among senior students (students of final year).

Predictive data mining is also used for support advisor in decision making. In this case, we were interested in finding impact of difference between the registered and gained credit hours on future learning behaviour using whole available data. for this purpose, several classification techniques with 10-fold Cross-validation were applied. Among of them, C4.5 constitutes the best agreement among the finding results.

**Author**

**Mohammed Al-Sarem:** Dr. Al-Sarem is an assistant professor of information science at the Taibah University, Al Madinah Al Monawarah, KSA. He received the PhD in Informatics from Hassan II University, Mohammadia, Morroco in 2014. His research interests center on E-learning, educational data mining, Arabic text mining, and intelligent and adaptive systems. He published several research papers and participated in several local/international conferences.

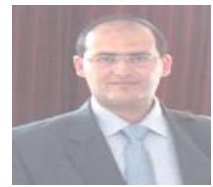